\documentclass[epj,referee]{svjour}
\usepackage{graphics}
\begin{document}
\title{Hamiltonian formalism of fractional systems}

\author{Aleksander~A.~Stanislavsky 
\thanks{e-mail: alexstan@ri.kharkov.ua}%
}                     
\institute{Institute of Radio Astronomy, 4 Chervonopraporna St.,
Kharkov 61002, Ukraine}
\date{}
%
\abstract{In this paper we consider a generalized classical
mechanics with fractional derivatives. The generalization is based
on the time-clock randomization of momenta and coordinates taken
from the conventional phase space. The fractional equations of
motion are derived using the Hamiltonian formalism. The approach
is illustrated with a simple-fractional oscillator in a free state
and under an external force. Besides the behavior of the coupled
fractional oscillators is analyzed. The natural extension of this
approach to continuous systems is stated. The interpretation of
the mechanics is discussed.
} 
\maketitle
\section{Introduction}\label{intro}
\thispagestyle{empty} The fractional differential equations become
very popular for describing anomalous transport,
diffusion-reaction processes, superslow relaxation, etc
\cite{1,2,2a,3} (and the references therein). The interest is
stimulated by the applications in various areas of physics,
chemistry and engineering \cite{3a,4,5,5a,6,6a,6b}. Nevertheless,
the derivation of such equations from some first physical
principles is not an easy matter. The fractional operator reflects
intrinsic dissipative processes that are sufficiently complicated
in nature. Their theoretical relationship with fractional calculus
is not yet ascertained in full.

The classical Hamiltonian (or Lagrangian) mechanics is formulated
in terms of derivatives of integer order. This technique suggests
advanced methods for the analysis of conservative systems, while
the physical world is rather nonconservative because of friction.
The account of frictional forces in physical models increases the
complexity in the mathematics needed to deal with them.

The fractional Hamiltonian (and Lagrangian) equations of motion
for the nonconservative systems were introduced into consideration
by Riewe \cite{7}. His approach is based on a simple look-out. If
the frictional force is proportional to velocity, the functional
form of a classical Lagrangian without friction may be added by a
term with the fractional derivative of one-half order. After
applying the variational technique, the obtained equation of
motion for the nonconservative system contains a contribution of
the frictional force. An extension of the fractional variational
problem for constrained systems was offered in \cite{7a,7b}.
However, as it was observed by Dreisigmeyer and Young~\cite{8},
the conception conflicts with the principle of causality. The
point is that the variational principle utilizes the integration
by part. After this operation the left Riemann-Louville fractional
derivative transforms into the right one which has a reversal
arrow of time. Unfortunately, the attempt to overcome this problem
by means of the treatment of the action as a Volterra series was
unsuccessful \cite{8a}.

An alternative approach for determining the fractional equations
of motion has been developed recently in \cite{9}. They are
derived by means of the fractional normalization condition. This
condition can be considered as a normalization condition for the
distribution function in a fractional space. The volume element of
the fractional phase space is realized by fractional exterior
derivatives. In this case the fractional system is described by a
fractional power of coordinates and momenta. Therefore, such
fractional systems are essentially nonlinear.

The purpose of the present paper is to provide a general method
for describing the fractional systems from some first principles.
We will look at the problem from another point of view. The main
difference of our consideration from above is a deep interrelation
of the fractional temporal derivative with stable distributions
from the theory of probability. Also, we intend to investigate
some concrete physical applications. Our paper is organized as
follows. The appearance of the fractional derivative in the
equations of motion is conditioned on a peculiar interaction of a
physical system with environment. In Sect.~\ref{sec:1} it is shown
that the interaction is taken into account through the temporal
variable that represents a sum of random intervals identically
distributed. The probability distribution of such a sum
asymptotically tends to a stable distribution. Using this approach
we derive the equation of motion in Sect.~\ref{sec:2}. Based on
the derivation the next section is devoted to the fractional
oscillator. The effect of the fractional damping on resonance is
considered in Sect.~\ref{sec:4}. This feature is similar to an
exponential damping, but there are some differences. Further the
dynamics of coupled fractional oscillators is represented in
Sect.~\ref{sec:5}. The forced oscillations of a multiple
fractional system are analyzed in Sect.~\ref{sec:6}. The
generalization of discrete fractional systems on a continuous case
is suggested in Sect.~\ref{sec:7}. In conclusion we discuss an
interpretation of the fractional mechanics.

\section{Operational time and subordination}\label{sec:1}
The spacetime is often treated as a continuum. It is something
like an ideal elastic medium. The conjecture is the overarching
principle of all physics. However, the concept is a fiction.
Nowhere does this feature of spacetime show itself more clearly
than at big bang and at collapse that it cannot be a continuum.
The ``elasticity'' of spacetime is only an approximation. There
are very different viewpoints to the space-time problem. One hopes
that an appropriate model for spacetime will be built in some
satisfactory theory of quantum gravitation. Others opt for some
type of discrete structures abandoning a continuum. In any case
the spacetime is too rich a structure to be pinned down by a
single description. Perhaps, several overlapping and possible
incompatible descriptions are need to exhaust the complex variety
of spacetime.

Usually, in the ordinary mechanical description, characterizing
the motion of a point particle, the time variable is
deterministic. Assume that the time variable represents a sum of
random temporal intervals $T_i$ being nonnegative independent and
identically distributed. Recall here the basic fact about the
density of a positive stable random variable. Following \cite{9a},
the density $g_\alpha(y)$ of such a variable is defined by its
characteristic function
\begin{displaymath}
\int_0^\infty e^{j\omega
y}\,g_\alpha(y)\,dy=\exp\{-|\omega|^\alpha\exp[-j\,(\pi\alpha/2)\,{\rm
sign}(\omega)]\}\,.
\end{displaymath}
If the waiting times $T_i$ belong to the strict domain of
attraction of an $\alpha$-stable distribution ($0<\alpha<1$),
their sum $n^{-1/\alpha}(T_1+T_2+\cdots+T_n),\, n\in\mathbf{N}$
converges in distribution to a stable law. The choice of the index
$\alpha$ in the range $0<\alpha<1$ is caused by the support of the
time steps $T_i$ on the nonnegative semiaxis. The continuous limit
of the discrete counting process $\{N_t\}_{t\geq
0}=\max\{n\in\mathbf{N}\mid \sum_{i=1}^nT_i\leq t\}$ is a first
passage time. The process is conventionally denoted by $S(t)$. For
a fixed time it represents the first passage of the stochastic
time evolution above that time level. The random process $S(t)$ is
non-decreasing and depends on the true time \cite{10}. As is well
known from the everyday experience, time is always running from
the past to the future. The important feature of time turns out to
be saved, if one chooses $S(t)$ as a new time clock. The random
process gives rise to the stochastic time arrow that is different
on the ordinary deterministic arrow \cite{11}.

Although the random process $S(t)$ is self-similar, it has neither
stationary nor independent increments, and all its moments are
finite \cite{12}. This process is non-Markovian, but it is inverse
to the continuous limit of a Markov random process of temporal
steps $T(\tau)$, i.\ e.\ $S(T(\tau))=\tau$. The probability
density of the process $S(t)$ has the following Laplace image
\begin{equation}
p^{S}(t,\tau)=\frac{1}{2\pi j}\int_{Br} e^{ut-\tau
u^\alpha}\,u^{\alpha-1}\,du\,, \label{eq21}
\end{equation}
where $Br$ denotes the Bromwich path, and $j=\sqrt{-1}$. This
probability density determines the probability to be at the
internal time $\tau$ on the real time $t$ \cite{11}. Carrying out
the change $ut\to u$ and denoting $z=\tau/t^\alpha$, the function
$p^{S}(t,\tau)$ is written as $t^{-\alpha}F_\alpha(z)$, where
\begin{displaymath}
F_\alpha(z)=\frac{1}{2\pi
j}\int_{Br}e^{u-zu^\alpha}\,u^{\alpha-1}\,du\,.
\end{displaymath}
This integral can be studied in almost exactly the same way as
Mainardi \cite{13} originally attacked the probability density
distribution for anomalous diffusion. To bend the Bromwich path
into the Hankel path, the function $F_\alpha(z)$ can be expanded
as a Taylor series. Besides, it has the Fox' H-function
representation \cite{13a}. Consequently, we have
\begin{displaymath}
F_\alpha(z)=H^{10}_{11}\left(z\Big|{(1-\alpha,\alpha)\atop
(0,1)}\right)=\sum_{k=0}^\infty\frac{(-z)^k}{k!
\Gamma(1-\alpha(1+k))}\,,
\end{displaymath}
where $\Gamma(x)$ is the usual gamma function.  The function
$F_\alpha(z)$ is entire and non-negative in $z>0$. It vanishes
exponentially for large positive $z$. Taking into account the
normalization relation $\int_0^\infty F_\alpha(z)\,dz=1$, it will
readily be seen that the function $p^{S}(t,\tau)$ is really a
probability density.

In the theory of anomalous diffusion the random process $S(t)$ is
used for the subordination of other random processes. Recall that
a subordinated process $Y(U(t))$ is obtained by randomizing the
time clock of a random process $Y(t)$ using a random process
$U(t)$ called the directing process.  The latter process is also
often referred to as the randomized time or operational time
\cite{13b}. These concepts are helpful in formulating the
continuous time random walk approach, deriving the equation of
subdiffusion, finding its solutions and moments, obtaining a
separation ansatz and in many others. The various features were
exhaustively analyzed in literature (see, e.\ g.,
\cite{13c,13d,14,14a,14b,14c,14cc,14d,14e}). The mathematically
elegant survey is represented in the paper \cite{12} that
especially is recommended for a deep insight into the
interrelation between fractional calculus and the theory of
probability. Nevertheless, it should be noticed that the exact
physical nature of the power-law waiting-time distributions in the
subordination scheme is unclear. The problem requires a more
rigorous examination elsewhere.

\section{Equations of motion}\label{sec:2}
Many important problems from classical mechanics can be solved
using the Newtonian formalism. Among them is the problem of motion
in a central force field which is of fundamental importance in
celestial mechanics. On the other hand the Lagrangian formalism is
more suited for handling problems in the theory of small
oscillations and in studying the dynamics of a rigid body.
Hamiltonian mechanics contains Lagrangian mechanics as a special
case. In addition the Hamiltonian formalism permits ones to solve
some problems (as the attraction by two stationary centers and the
determination of geodesics on the triaxial ellipsoid) which do not
yield solutions by other means. It is expected that this route may
be followed in the analysis of fractional systems.

Let a Hamiltonian system evolution depend on operational time
$\tau$. The corresponding equation of motion is written as
\begin{equation}
\frac{dq}{d\tau}=\frac{\partial\mathcal{H}}{\partial
p}\,,\qquad\qquad\frac{dp}{d\tau}=-\frac{\partial
\mathcal{H}}{\partial q}\,.\label{eq31}
\end{equation}
Consider a dynamical system for which the momentum and the
coordinate satisfy the relations
\begin{eqnarray}
p_\alpha(t)=\int_0^\infty
p^{S}(t,\tau)\,p(\tau)\,d\tau\,,\nonumber\\
q_\alpha(t)=\int_0^\infty
p^{S}(t,\tau)\,q(\tau)\,d\tau\,.\nonumber
\end{eqnarray}
Their Laplace transform with respect to time has a simple
algebraic form
\begin{eqnarray}
\bar p_\alpha(s)=\int_0^\infty
e^{-st}p_\alpha(t)\,dt=s^{\alpha-1}\bar p(s^\alpha)\,,\nonumber\\
\bar q_\alpha(s)=\int_0^\infty
e^{-st}q_\alpha(t)\,dt=s^{\alpha-1}\bar q(s^\alpha)\,,\nonumber
\end{eqnarray}
where $\alpha$ is in the range of $0<\alpha<1$. Since the values
$\partial\mathcal{H}/\partial p$\ \ and\ \
$\partial\mathcal{H}/\partial q$\ \ depend on operational time, we
assume that
\begin{eqnarray}
\frac{\partial\mathcal{H}_\alpha}{\partial p_\alpha}=\int_0^\infty
p^{S}(t,\tau)\,\frac{\partial\mathcal{H}}{\partial
p}\,d\tau\,,\nonumber\\
\frac{\partial\mathcal{H}_\alpha}{\partial q_\alpha}=\int_0^\infty
p^{S}(t,\tau)\,\frac{\partial\mathcal{H}}{\partial
q}\,d\tau\,.\nonumber
\end{eqnarray}
In this case Eqs.(\ref{eq31}) become fractional:
\begin{equation}
\frac{d^{\alpha}q_\alpha}{d
t^{\alpha}}=\frac{\partial\mathcal{H}_\alpha}{\partial
p_\alpha}=p_\alpha\,,\qquad\frac{d^{\alpha}p_\alpha}{d
t^{\alpha}}=-\frac{\partial\mathcal{H}_\alpha}{\partial
q_\alpha}\,.\label{eq32}
\end{equation}
Here we use the so-called Caputo derivative \cite{14aa,14ab,14ac}
defined as
\begin{eqnarray}
&&\frac{d^{\alpha}x(t)}{d t^{\alpha}}=\tilde D^\alpha
x(t)=J^{n-\alpha}D^nx(t)=\nonumber\\
&&=\frac{1}{\Gamma(n-\alpha)}\int^t_0\frac{x^{(n)}(\tau)}
{(t-\tau)^{\alpha+1-n}}\,d\tau,\quad n-1<\alpha<n,\nonumber
\end{eqnarray}
where $x^{(n)}(t)=D^nx(t)$ means the $n$-derivative of $x(t)$, and
\begin{displaymath}
J^\beta y(t)=\frac{1}{\Gamma(\beta)}\int^t_0
y(\tau)\,(t-\tau)^{\beta-1}\,d\tau
\end{displaymath}
is the fractional integral. In fact, the power-law waiting times
imply fractional derivatives in time. The reader who desires
further background information on fractional calculus will do well
to consult the excellent books \cite{14ad,14ae}.

When $\alpha=1$, the generalized equations (\ref{eq32}) transform
into the ordinary, namely Hamiltonian equations being well known
from classical mechanics. Thus this point of view embraces a wide
circle of physical tasks. In next sections we will elaborate some
of them.

\section{Fractional oscillator}\label{sec:3}
\begin{figure*}
\centering
\resizebox{1.6\columnwidth}{!}{%
  \includegraphics{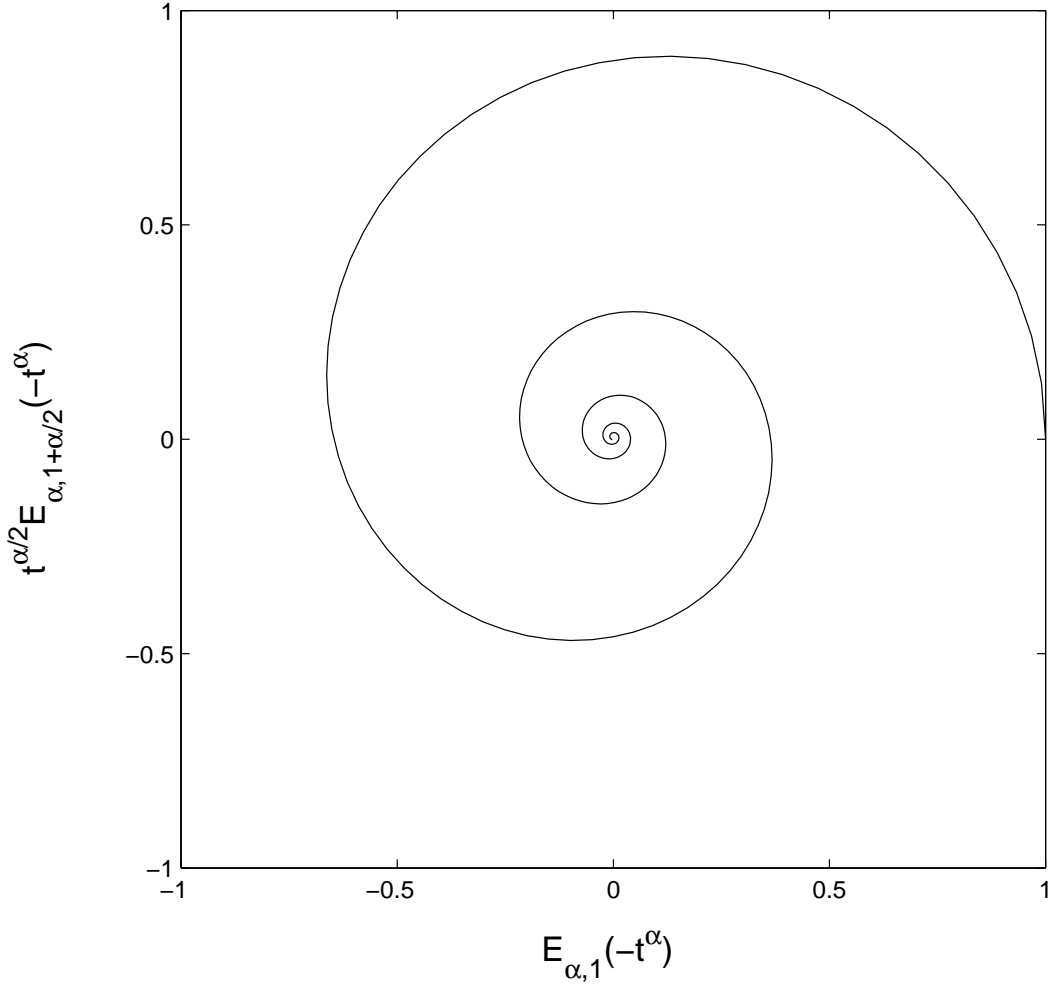}
}  \caption{Phase plane diagram of the fractional oscillator with
the initial conditions $q_\alpha(0)=1$, $\tilde D^{\alpha}
q_\alpha(0)=0$, where $\alpha=1.8$.} \label{fig:1}
\end{figure*}

One of the simplest  physical models (but nontrivial) supported by
the above-mentioned method is a fractional oscillator. Its
generalized Hamiltonian takes the form
\begin{equation}
H_\alpha=(p^2_\alpha+\omega^2q^2_\alpha)/2\,,\label{eq41}
\end{equation}
where $\omega$ is the circular frequency, $q_\alpha$ and
$p_\alpha$ the displacement and the momentum respectively. The
value describes the total energy of this dynamical system
\cite{15}. Although the Hamiltonian (\ref{eq41}) is not an
explicit function of time, for non-integer values $\alpha$ the
system is nonconservative because of the fractional derivative of
momentum. Then the Hamiltonian equations for the fractional
oscillator are written as
\begin{eqnarray}
\tilde D^{\alpha}q_\alpha&=&\frac{d^{\alpha}q_\alpha}{d
t^{\alpha}}=\frac{\partial H_\alpha}{\partial
p_\alpha}=p_\alpha\,,\label{eq42}\\ \tilde
D^{\alpha}p_\alpha&=&\frac{d^{\alpha}p_\alpha}{d
t^{\alpha}}=-\frac{\partial H_\alpha}{\partial
q_\alpha}=-\omega^2q_\alpha\,.\label{eq43}
\end{eqnarray}
It follows from this that
\begin{equation}
\tilde D^{2\alpha}q_\alpha+\omega^2q_\alpha=0\,\qquad\mathrm
{or}\qquad  \tilde D^{2\alpha}
p_\alpha+\omega^2p_\alpha=0.\label{eq44}
\end{equation}
Each of the equations has two independent solutions. Suffice it to
solve one of these equations, for example, that determines the
coordinate:
\begin{equation}
q_\alpha(t)=A\,E_{2\alpha,1}(-\omega^2t^{2\alpha})+B\,\omega
t^{\alpha}
E_{2\alpha,\,1+\alpha}(-\omega^2t^{2\alpha})\,,\label{eq8}
\end{equation}
where $A$, $B$ are constants, and
\begin{displaymath}
E_{\mu,\,\nu}(z)=\sum_{k=0}^{\infty}\frac{z^k}{\Gamma(\mu k+\nu)},
\qquad \mu,\nu>0,
\end{displaymath}
is the two-parameter Mittag-Leffler function. Then the fractional
oscillator momentum is expressed in terms of the fractional
derivative of the coordinate
\begin{displaymath}
p_\alpha(t)=m\,\tilde D^{\alpha} q_\alpha(t)\,,
\end{displaymath}
where $m$ is the generalized mass \cite{15}. In this connection it
is relevant to remark that
\begin{eqnarray}
&\tilde D^{\alpha}&E_{2\alpha,\,1}(-t^{2\alpha})=-t^{\alpha}
E_{2\alpha,\,1+\alpha}(-t^{2\alpha})\,,\nonumber\\ &\tilde
D^{\alpha}&t^{\alpha}E_{2\alpha,\,1+\alpha}(-t^{2\alpha})=
E_{2\alpha,\,1}(-t^{2\alpha})\,.\nonumber
\end{eqnarray}
The phase portrait is a close curve only for the harmonic
oscillator ($2\alpha=2$), but in our case this is a spiral. The
total energy decreases. An example of the phase plane diagram for
the fractional oscillator is represented in Fig.~\ref{fig:1}. The
intrinsic dissipation in the fractional oscillator is cau-sed by
the following reason. The fractional oscillator may be considered
as an ensemble average of harmonic oscillators \cite{15a}. The
oscillators differ slightly from each other in frequency because
of the subordination. Therefore, even if they start in phase,
after a time the oscillators will be allocated uniformly up to the
clock-face. Although each oscillator is conservative, the system
of such oscillators with the dynamics like a fractional oscillator
demonstrates a dissipative process stochastic by nature.

\begin{figure*}
\centering
\resizebox{1.6\columnwidth}{!}{%
  \includegraphics{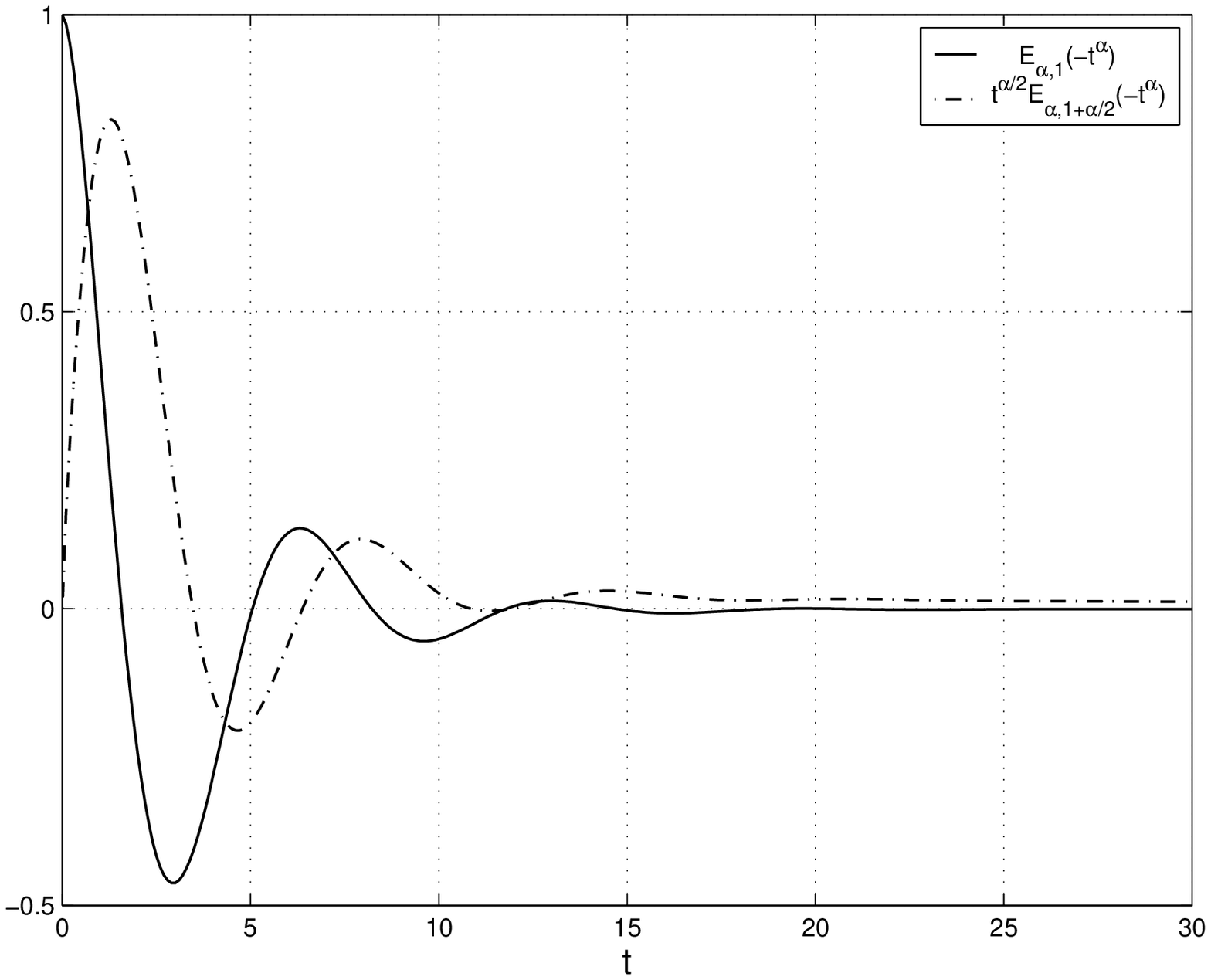}
} \caption{Fractional oscillations are characterized by a finite
number of zeros ($\alpha=1.65$).} \label{fig:11}
\end{figure*}


As is shown in \cite{14ac,16}, the fractional oscillations exhibit
a finite number of damped oscillations with an algebraic decay
(Fig.~\ref{fig:11}). The fact rests on that such oscillations may
be decomposed into two parts. One of them demonstrates
asymptotically an algebraic (monotonic) decay, and another term
represents an exponentially damped harmonic oscillation. Owing to
the second term decreasing faster than this happens for the term
with an algebraic decay, the fractional oscillations possess a
finite number of zeros. Our notation $\omega_0$ named as a
circular frequency is enough acceptable, since the value
characterizes the number of oscillations on a given temporal
interval.

\section{Effect of fractional damping on resonance}\label{sec:4}
Now we consider a behavior of the fractional oscillator under an
external force. Following the initial conditions $x(0)=0$ and
$\dot x(0)=0$, this model is described by the following equation
\begin{eqnarray}
x(t)&=&-\frac{\omega^\alpha_0}{\Gamma(\alpha)}\int^t_0(t-t')^{\alpha-1}\,
x(t')\,dt'+\nonumber\\
&+&\frac{1}{\Gamma(\alpha)}\int^t_0(t-t')^{\alpha-1}\,
F(t')\,dt'\,,\label{eq51}
\end{eqnarray}
where it should be kept $1<\alpha<2$, and $F$ is the external
force. The dynamic response of the driven fractional oscillator
was investigated in \cite{15}:
\begin{equation}
x(t)=\int^t_0F(t')\,(t-t')^{\alpha-1}\,E_{\alpha,\,\alpha}
(-\omega^\alpha_0(t-t')^\alpha)\,dt'\,. \label{eq52}
\end{equation}
This allows us to define the response for any desired forcing
function $F(t)$. The ``free'' and ``forced'' oscillations of such
a fractional oscillator depend on the index $\alpha$. However, in
the first case the damping is characterized only by the ``natural
frequency'' $\omega_0$, whereas the damping in the case of
``forced'' oscillations depends also on the driving frequency
$\omega$. Each of these cases has a characteristic algebraic tail
associated with damping \cite{16a}.

\begin{figure*}
\centering
\resizebox{1.6\columnwidth}{!}{%
  \includegraphics{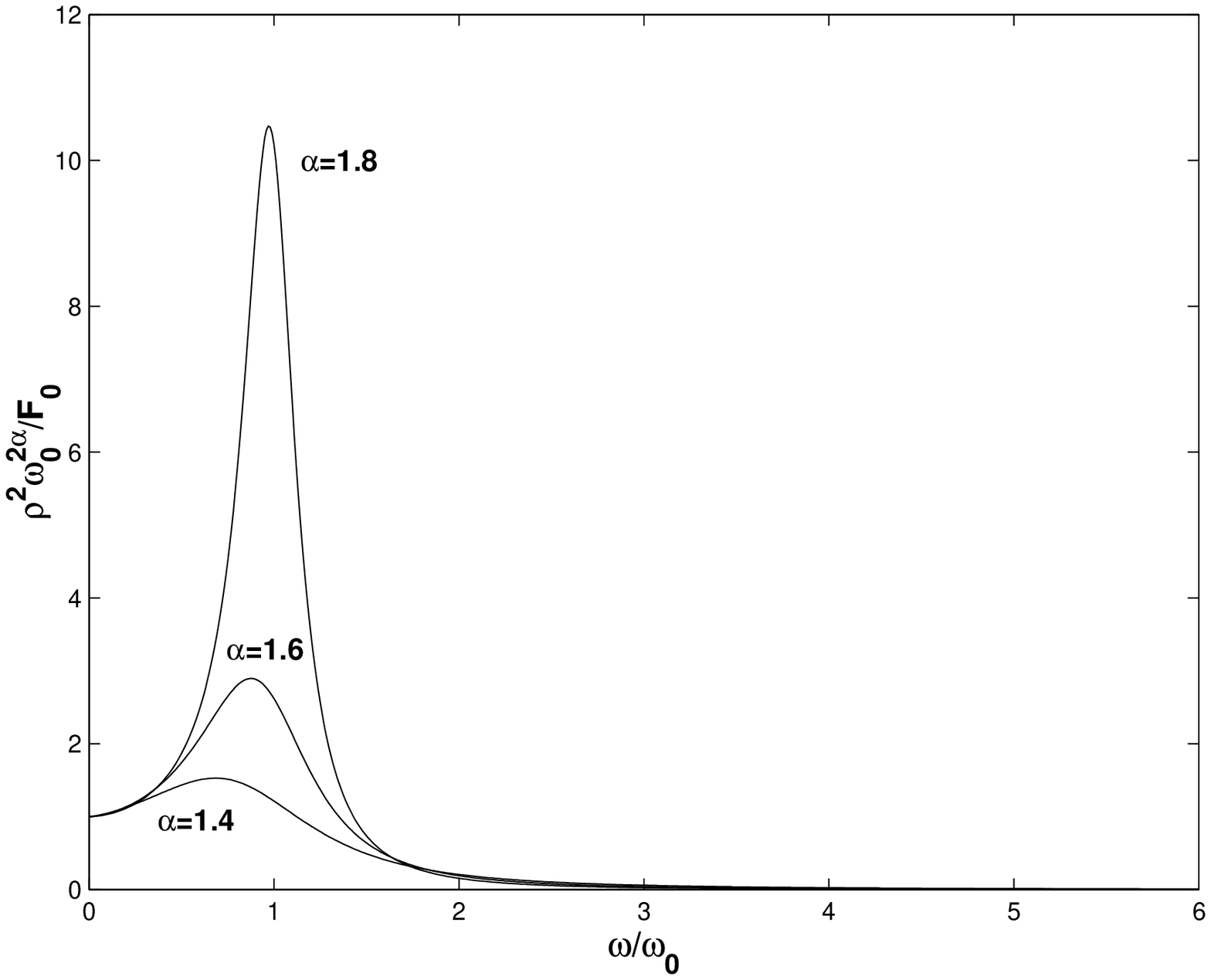}
}\\
\resizebox{1.6\columnwidth}{!}{%
  \includegraphics{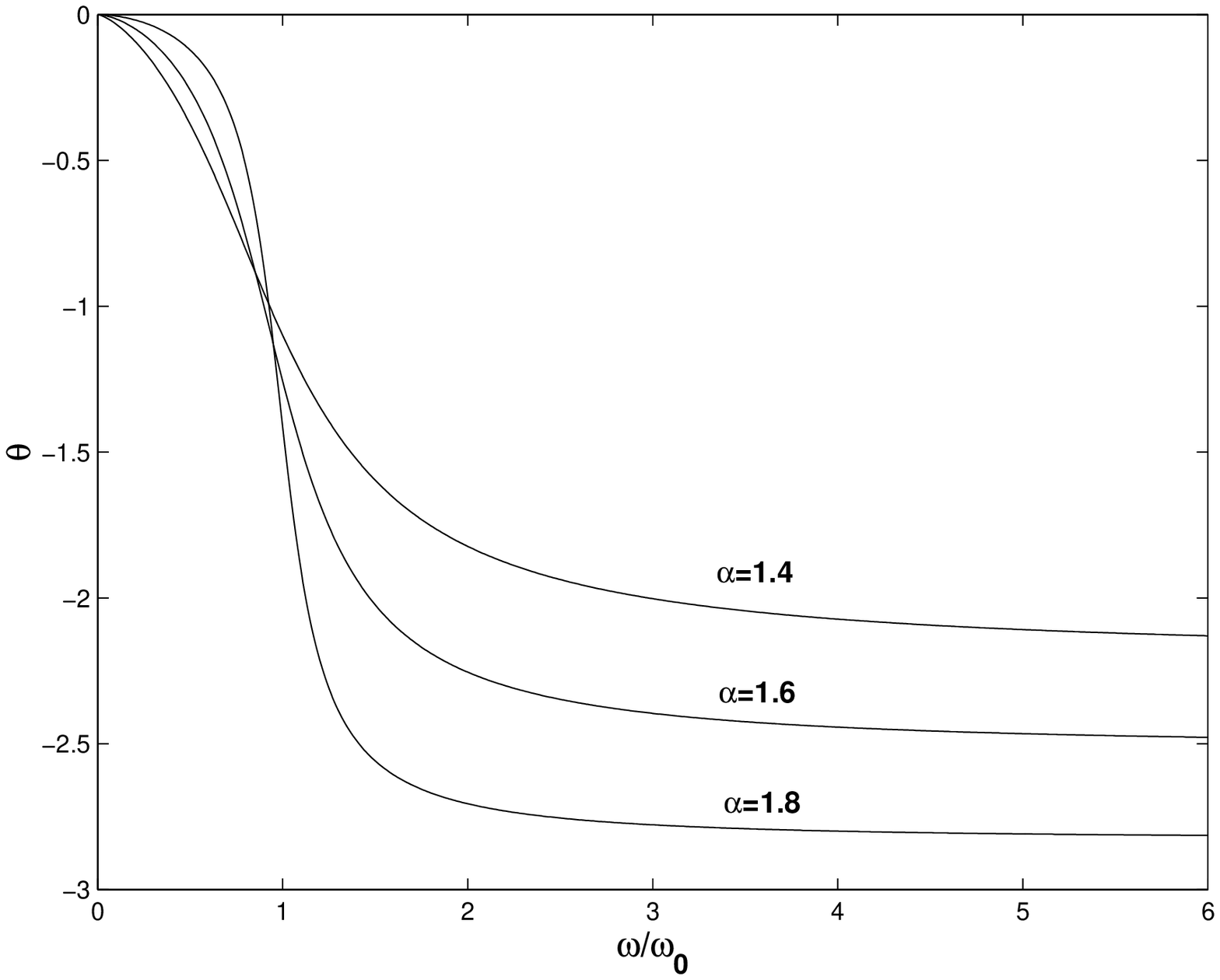}
} \caption{Resonance curves and phase shift of the fractional
oscillator forced by an harmonic oscillation with the frequency
$\omega$.} \label{fig:1ab}
\end{figure*}


If $F(t)$ is periodic, namely $F(t)=F_0\,e^{j\omega t}$, then the
solution of Eq. (\ref{eq51}) is determined by taking the inverse
Laplace transform
\begin{equation}
x(t)=\frac{1}{2\pi
j}\int_{Br}e^{st}\frac{F_0\,(s+j\omega)\,ds}{(s^2+\omega^2)\,
(s^\alpha+\omega_0^\alpha)}\,.\label{eq53}
\end{equation}
The Bromwich integral (\ref{eq53}) can be evaluated in terms of
the theory of complex variables. Some particular examples of
driving functions in this context were considered in \cite{15}. However, the case
under interest was studied in \cite{15a}. Consider only the
harmonically forced oscillation. If one waits for a long enough
time, the normal mode in this system is damped. After the
substitution $x_0e^{j\omega t}$ for $x(t)$ in Eq.(\ref{eq51}) we
obtain
\begin{eqnarray}
x_0\,e^{j\omega t}
&=&-\frac{\omega^\alpha_0}{\Gamma(\alpha)}\int^t_0(t-t')^{\alpha-1}\,
x_0\,e^{j\omega t'}\,dt'+\nonumber\\
&+&\frac{1}{\Gamma(\alpha)}\int^t_0(t-t')^{\alpha-1}\,
F_0\,e^{j\omega t'}\,dt'\,.\label{eq54}
\end{eqnarray}
It is convenient to change the variable $\omega(t-t')=\zeta$ in
the integrand. Next we can divide out $\exp(j\omega t)$ from each
side of (\ref{eq54}) and direct $t$ to infinity. The procedure
permits one to extract the contribution of steady-state
oscillations. Then Eq. (\ref{eq54}) gives
\begin{equation}
x_0=\frac{F_0}{[\omega_0^\alpha+\omega^\alpha\exp(j\pi\alpha/2)]}
\,.\label{eq55}
\end{equation}
The forced solution $\rho\exp(j\omega t+\theta)$ is written as
\begin{eqnarray}
&\rho^2&=\frac{F_0}{[\omega_0^{2\alpha}+\omega^{2\alpha}+
2\omega_0^\alpha\omega^\alpha\cos(\pi\alpha/2)]}\,,\nonumber\\
&\tan\theta&=-\frac{\omega^\alpha\sin(\pi\alpha/2)}{\omega_0^\alpha+
\omega^\alpha\cos(\pi\alpha/2)}\,.\nonumber
\end{eqnarray}
Denote $\omega/\omega_0=z$. Then we have
\begin{eqnarray}
\frac{\rho^2\omega_0^{2\alpha}}{F_0}&=&\frac{1}{[z^{2\alpha}+
2z^\alpha\cos(\pi\alpha/2)+1]}\,,\nonumber\\
\tan\theta&=&-\frac{z^\alpha\sin(\pi\alpha/2)}{1+
z^\alpha\cos(\pi\alpha/2)}\,.\label{eq56}
\end{eqnarray}
It should be noticed that the maximum of
$\rho^2\omega_0^{2\alpha}/F_0$ is not attained for
$\omega=\omega_0$. It is shifted to the origin of coordinates
on Fig.~\ref{fig:1ab} with decaying $\alpha$. To differentiate
$\rho^2\omega_0^{2\alpha}/F_0$ with respect to $z$, we get the
maximum $z_{max}=\sqrt[\alpha]{\cos[\pi(2-\alpha)/2]}$. For
$\alpha=2$ the damping in this oscillator vanishes. If then both
frequencies coincide $\omega=\omega_0$, the amplitude of the
oscillator tends to infinity.

\section{Coupled fractional oscillators}\label{sec:5}
From the theory of vibrations it is well known, much of physical
systems permits a description in the form of free harmonic
oscillators. However, such a representation is too idealized, as
in many practical cases these systems are not usually isolated.
Instead, they interact with environment, or with other
oscillators. Therefore, the study of the dynamics of driven and
coupled oscillators is of major importance.

Here we intend to provide a similar analysis for fractional
systems. Consider two identical fractional oscillators mutually
coupled. For $1<\alpha\leq 2$ the dynamics of this system is given
by
\begin{eqnarray}
\tilde D^\alpha
x_1+\omega_0^2x_1+\kappa^2(x_1-x_2)&=&0\,,\label{eq61}\\ \tilde
D^\alpha x_2+\omega_0^2x_2+\kappa^2(x_2-x_1)&=&0\,,\label{eq62}
\end{eqnarray}
where the two oscillators are labelled by 1 and 2, respectively,
and $\kappa^2$ is the measure of the coupling, $\omega_0$ the
circular frequency. Let us introduce new variables $u_1=x_1+x_2$
and $u_2=x_1-x_2$. In such coordinates the system of equations
(\ref{eq61}), (\ref{eq62}) transforms into the equations of two
independent fractional oscillators with the frequencies $\omega_0$
and $\sqrt{\omega_0^2+2\kappa^2}$. If $u_2=0$, then $x_1=x_2$ so
that both oscillators move in phase with the frequency $\omega_0$.
In this case the coupling between the oscillators has no influence
on their motion. If $u_1=0$ or $x_1=-x_2$ that is the same, the
oscillators evolve in antiphase with the increased frequency
$\sqrt{\omega_0^2+2\kappa^2}$ in force of the measure of the
coupling.

It is often found that the physical systems of interest consist of
two and more oscillators which interact weekly among them. When
the coupling is weak ($\kappa\ll 1$), the fractional oscillators
transfer their energy from ones to others and vice versa. The
effect depends on the magnitude of the index $\alpha$
characterizing a strength of dissipation as well as on the
magnitude of the parameter $\kappa$. Nevertheless, due to the
dissipation, finally the fractional oscillations will decrease in
amplitude.

\begin{figure*}
\centering
\resizebox{1.6\columnwidth}{!}{%
  \includegraphics{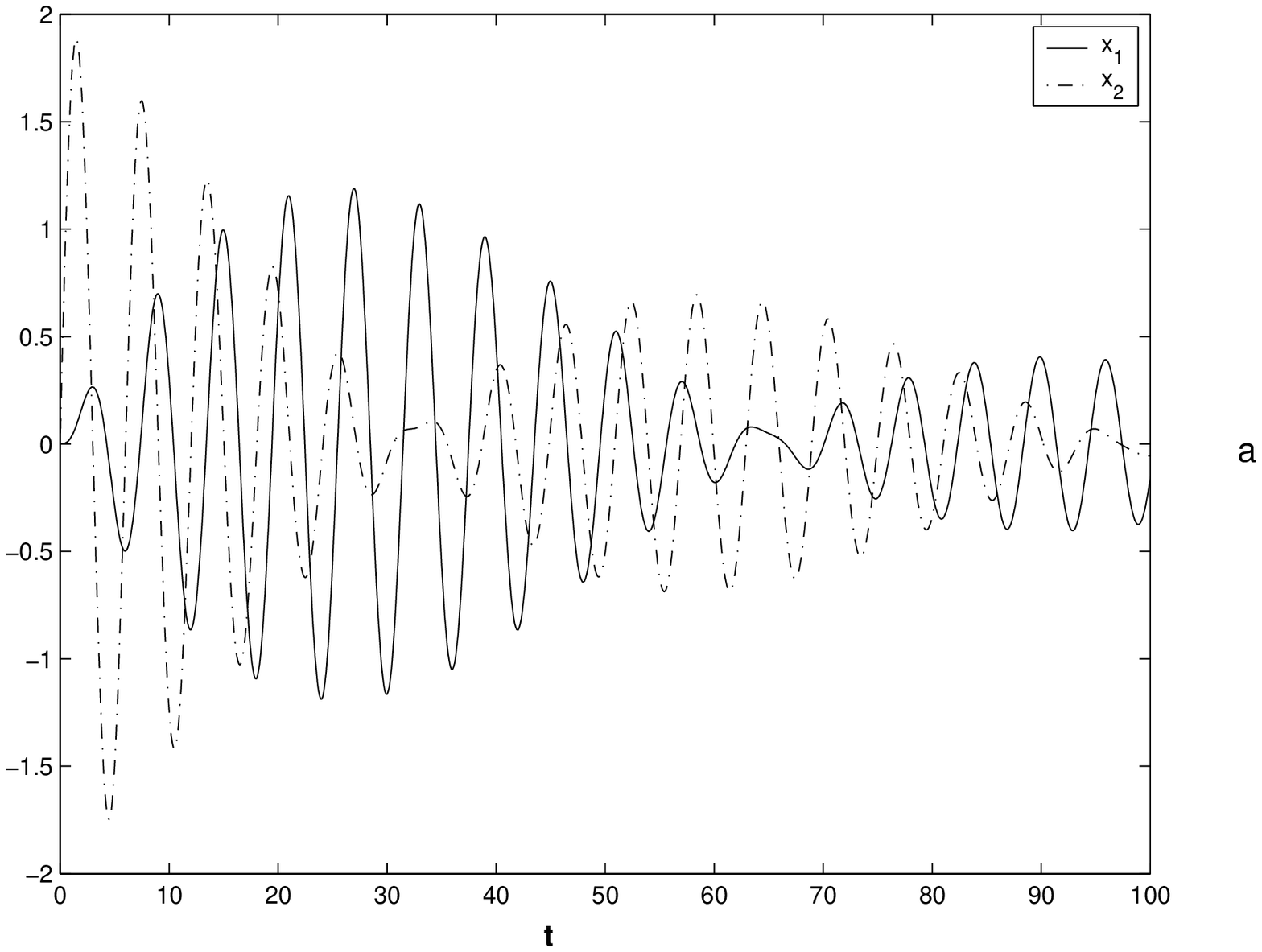}
}
\resizebox{1.6\columnwidth}{!}{%
  \includegraphics{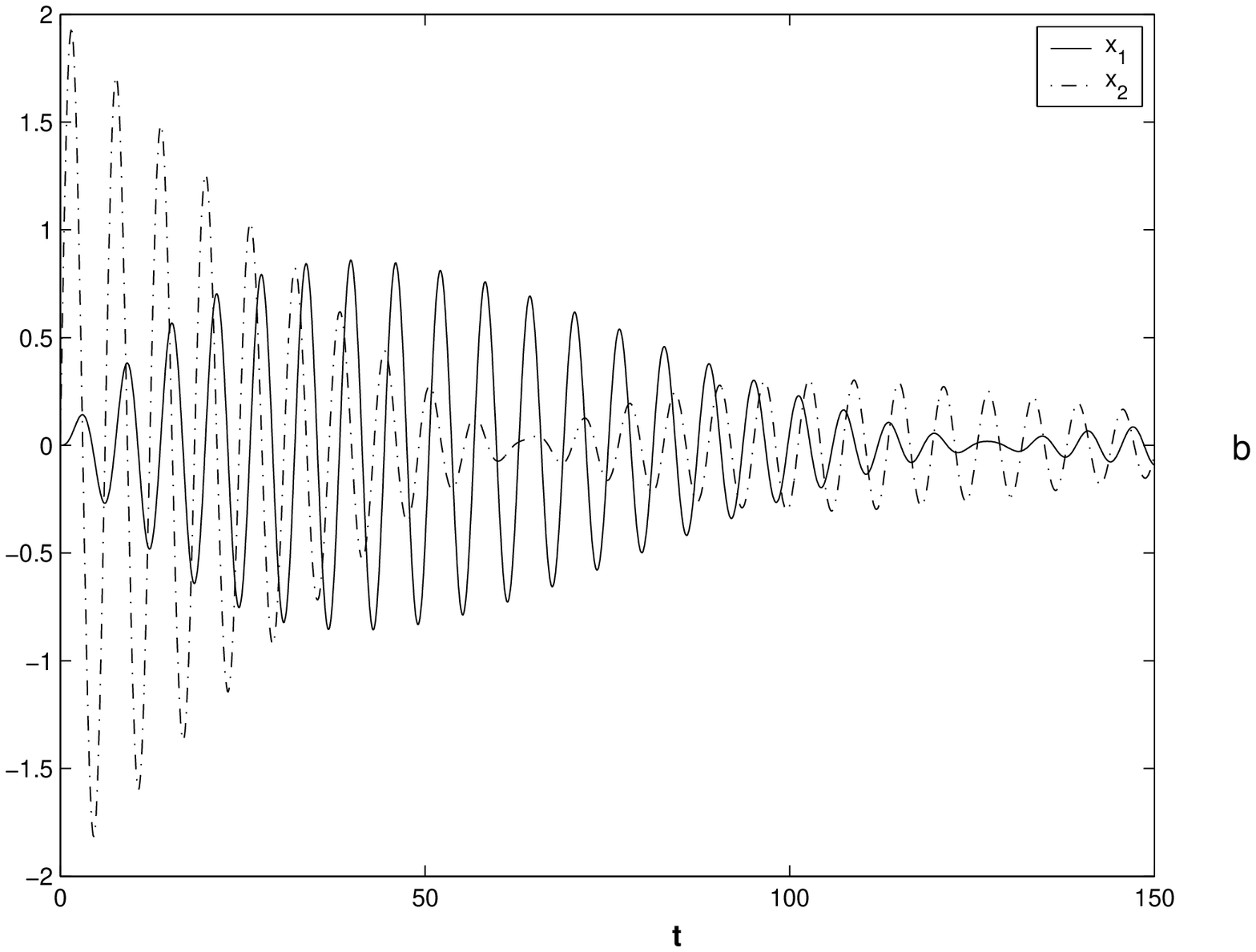}
}
\caption{Superposition of two modes ($x_1$ and $x_2$) with
$\alpha=1.98, \omega_0=1$: a) $\kappa=0.3162$; b)
$\kappa=0.2236$.}
\label{fig:2}       
\end{figure*}

Let two oscillators rest initially. Then one of them gets $\tilde
D^{\alpha/2} x_1(0)=B_0$. Following Sect.~\ref{sec:3}, the time
evolution of this system takes the form
\begin{eqnarray}
x_1(t)=\frac{B_0}{2}\Bigl[\omega_0 t^{\alpha/2}
E_{\alpha,\,1+\alpha/2}(-\omega^2_0 t^{\alpha})+\nonumber\\
+\,t^{\alpha/2} E_{\alpha,\,1+\alpha/2}(-(\omega_0^2+2\kappa^2)
t^{\alpha})\Bigr]\,,\label{eq63}\\
x_2(t)=\frac{B_0}{2}\Bigl[\omega_0 t^{\alpha/2}
E_{\alpha,\,1+\alpha/2}(-\omega^2_0
t^{\alpha})-\nonumber\\-\,t^{\alpha/2}
E_{\alpha,\,1+\alpha/2}(-(\omega_0^2+2\kappa^2)
t^{\alpha})\Bigr]\,.\label{eq64}
\end{eqnarray}
Fig.~\ref{fig:2} shows a superposition of two fractional
oscillations. The oscillations observable are complicated, but
they can be likened to damping beats. In the force of this damping
the minimum of $x_1$ does not coincide with the maximum of $x_2$.
The phase shift between them is determined by the value $\alpha$,
i.\ e.\ by the level of dissipation in this system. Moreover, the
algebraic relaxation term of the fractional oscillations deforms
the position of the maxima of $x_1$ and $x_2$. Distinctly this can
be seen in Fig.~\ref{fig:2}b\ .

The decomposition of the dynamics of coupled fractional
oscillators in a superposition of normal modes,\,\,\, which are
simple fractional oscillators, is a major result in the physics of
coupled fractional oscillators. This decomposition is not
restricted to just two coupled oscillators. An analogous
decomposition holds for an arbitrary large number of coupled
fractional oscillators.

\section{Forced oscillations of a multiple fractional system}\label{sec:6}
Let a fractional oscillator have two degree of freedom. The
harmonic force with the frequency $p$ acts on the system
coordinates $x$ and $y$. Then the equations of motion are written
as
\begin{equation}
\cases{A\tilde D^\alpha x+H\tilde D^\alpha y+ax+hy=X\cos pt\,,\cr
H\tilde D^\alpha x+B\tilde D^\alpha y+hx+by=Y\cos
pt\cr}\label{eq71}
\end{equation}
with $1<\alpha\leq 2$. We shall seek solutions to Eqs.
(\ref{eq71}) in the form
\begin{displaymath}
x=\beta\cos pt\,,\qquad\qquad y=\gamma\cos pt\,.
\end{displaymath}
The treatment of dynamical equations can be greatly simplified by
using the representation in complex numbers. With this object in
view, replace $\cos pt$ by $e^{jpt}$ in (\ref{eq71}). Next we
substitute $x=\beta e^{jpt}$ and $y=\gamma e^{jpt}$ in such a
system of equations. Pick out $e^{jpt}$ in each term of the
equations. The fractional derivative of $e^{jpt}$ gives
\begin{eqnarray}
\tilde D^\alpha e^{jpt}&=&J^{2-\alpha}D^2 e^{jpt}=
-p^2J^{2-\alpha}e^{jpt}=\nonumber\\ &=&-\frac{p^\alpha
e^{jpt}}{\Gamma(m)}\int_0^{pt}\tau^{m-1}\,e^{-j\tau}\,d\tau\,,
\nonumber
\end{eqnarray}
where $m=2-\alpha$ is constant. Making $t\to\infty$ in the above
integral, we apply the table integral \cite{17}:
\begin{displaymath}
\int_0^\infty z^{\xi-1}\,e^{-jz}\,dz=\Gamma(\xi)\,e^{-j\pi\xi/2}
\end{displaymath}
for $0<{\rm Re}\,\xi<1$. This technique transforms Eqs.
(\ref{eq71}) into the algebraic expressions
\begin{equation}
\cases{(a+A\,p^\alpha e^{j\pi\alpha/2})\beta+(h+H\,p^\alpha
e^{j\pi\alpha/2})\gamma=X\,,\cr (h+H\,p^\alpha
e^{j\pi\alpha/2})\beta +(b+B\,p^\alpha
e^{j\pi\alpha/2})\gamma=Y\,.\cr}\label{eq72}
\end{equation}
The steady-state solutions of Eqs. (\ref{eq71}) can be derived
from (\ref{eq72}) by taking the ``real'' part of $\beta$ and
$\gamma$, i.\ e.\ by projecting the motion onto the real axis of
$x$ and $y$. From Eqs.(\ref{eq72}) at once it follows that
\begin{eqnarray}
\beta=\frac{1}{\Delta}\left|
    \begin{array}{lllll}
       X\ \ & h+H\,p^\alpha e^{j\pi\alpha/2} \\
       Y\ \ & b+B\,p^\alpha e^{j\pi\alpha/2} \\
    \end{array}
       \right|\,,\nonumber\\
\gamma=\frac{1}{\Delta}\left|
    \begin{array}{lllll}
         a+A\,p^\alpha e^{j\pi\alpha/2}\ \ & X \\
         h+H\,p^\alpha e^{j\pi\alpha/2}\ \ & Y\\
    \end{array}
\right|\,,\label{eq73}
\end{eqnarray}
where
\begin{displaymath}
\Delta=\left|
  \begin{array}{lllll}
    a+A\,p^\alpha e^{j\pi\alpha/2}\ \ & h+H\,p^\alpha e^{j\pi\alpha/2}\\
    h+H\,p^\alpha e^{j\pi\alpha/2}\ \ & b+B\, p^\alpha e^{j\pi\alpha/2}\\
  \end{array}
\right|
\end{displaymath}
is the determinant. Thus the general conclusion is the following.
When a periodic force of simple-harmonic type acts on any part of
the system, every part executes a simple-harmonic vibration of the
same period, but the amplitude will be different in different
parts. When the period of the forced vibration nearly coincides
with one of the free modes, an abnormal amplitude of forced
vibration will be in general result, owing to the smallness of the
denominator on the formulae (\ref{eq73}).

If $X\neq 0$, $Y=0$, then
\begin{equation}
\gamma=-\frac{X(h+H\,p^\alpha
e^{j\pi\alpha/2})}{\Delta}\,.\label{eq74}
\end{equation}
Imagine now a second case of forced vibration in which $X=0$,
$Y\neq 0$. This yields
\begin{displaymath}
\beta=-\frac{Y(h+H\,p^\alpha e^{j\pi\alpha/2})}{\Delta}\,.
\end{displaymath}
Comparing with (\ref{eq74}), we see that
\begin{equation}
\gamma :X=\beta : Y\,.\label{eq75}
\end{equation}
The result concerns a remarkable theorem of reciprocity, first
proved for the theory of aerials by Helmholtz, and afterwards
greatly extended by Rayleigh. Their interpretation is most easily
expressed when the ``forces'' $X$, $Y$ are of the same character
in such a way that we may put $X=Y$ and obtain $\beta=\gamma$. The
coupled fractional oscillators forced by harmonic oscillations
support well the theorem. Indeed, it is clear in view of their
equations of motion being linear.

\begin{figure*}
\centering
  \includegraphics{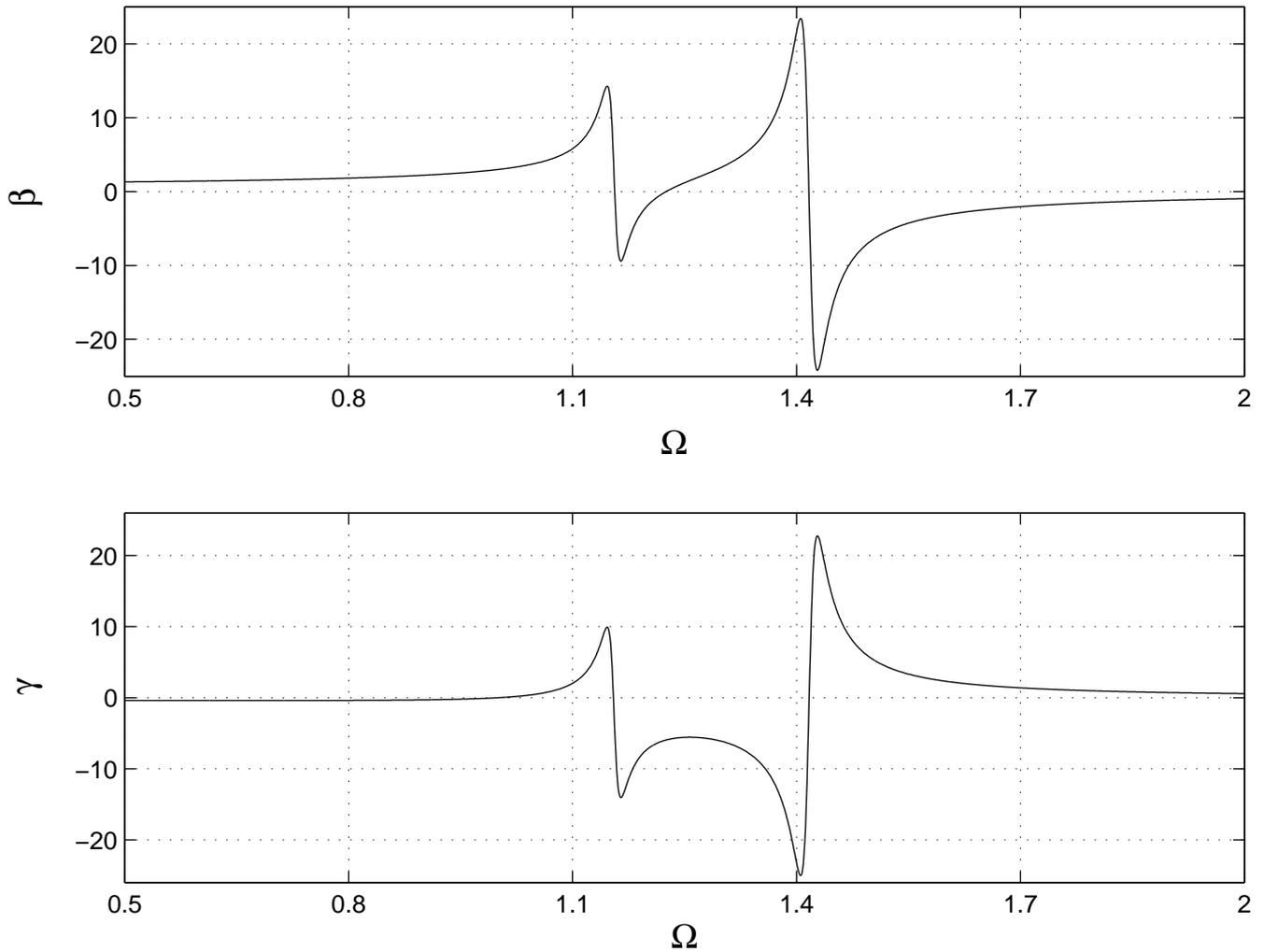}
\caption{Resonance curves of the fractional oscillator forced with
two degree of freedom: $X=3$, $Y=0$, $\alpha=1.99, H=h=1, A=B=2,
a=b=3$. The dynamical damping is detectable, when $\beta=0$ and
$\gamma\neq 0$. See text for more details.}
\label{fig:3}       
\end{figure*}

The resonance curves are represented in Fig.~\ref{fig:3}. Here we
consider the case, when $X\neq 0$, $Y=0$. The curves demostrate
the following interesting effects: 1) if the exterior force
frequency $\Omega$ coincides with one of normal frequencies of
this system, the amplitudes of both oscillators increase; 2) if
the frequency of the exterior force, acting on the first
oscillator, coincides with
$\sqrt[\alpha]{b/\{B\cos(\pi(2-\alpha)/2)\}}>0$, then the first
oscillator is rest ($\beta=0$). The latter phenomenon is called
the dynamic damping often used for the breaking of unfavourable
oscillations \cite{18}. For the exterior force frequency
$\sqrt[\alpha]{h/\{H\cos(\pi(2-\alpha)/2)\}}>0$ the second
oscillator will be rest.

The extension of the method to the general case of linear
fractional oscillations of a multiple system is obvious so that
the result may be stated formally. In such a system with $k$
degrees of freedom there are in general $k$ distinct ``normal
modes'' of free fractional vibrations about a configuration of
stable equilibrium, the frequencies of which are given by a
symmetrical determinantal equation of the $k$-th order in
$p^\alpha e^{j\pi\alpha/2}$, analogous to (\ref{eq72}). In each of
these modes the system oscillates exactly as if it had only one
degree of freedom.

\section{Transition to continuous systems}\label{sec:7}
At least mathematically, it is sometimes possible to pass from the
study of oscillatory systems of finite freedom to continuous
systems by a sort of limiting process. So D.~Bernoulli (1732)
considered the vibrations of a hanging chain as a limiting form of
the problem where a large number of equal and equidistant
particles are attached to a tense string whose own mass is
neglected. The general principles may be applied to fractional
systems too. In this section we shall be concerned with fractional
systems for which the number of degrees of freedom becomes
infinite.

The lattice represents the most illustrative example which is
naturally called by the ordered structure of oscillators. To start
with an one-dimensional chain of identical fractional oscillators,
the characteristic spatial period of wave motion in this chain is
assumed to be too more than the mesh dimension. If the oscillators
interact with the nearest neighbours, they are described by the
equations of type
\begin{equation}
\frac{d^\alpha\phi_n}{dt^\alpha}+\omega_0^2\phi_n=
M(\phi_{n-1}+\phi_{n+1}-2\phi_n)\,,\label{eq81}
\end{equation}
where $M$ is constant, and $1<\alpha\leq 2$. Proceeding from the
discrete system to a continuous one, we arrive at the equation in
partial derivatives
\begin{equation}
\frac{\partial^\alpha\phi}{\partial
t^\alpha}-v^2\,\frac{\partial^2\phi}{\partial
x^2}+\omega^2_0\phi=0\,, \label{eq82}
\end{equation}
where $v^2=Ma^2$, and $a$ is the mesh size. If $\omega_0=0$, the
equation (\ref{eq82}) is simplified so that its solution is
\begin{equation}
\phi(x,t)=\int_0^\infty
p^{S}(t,\tau)\,\Bigl\{f_1(v\tau-x)+f_2(v\tau-x)\Bigr\}\,d\tau,
\label{eq83}
\end{equation}
where the functions $f_1$, $f_2$ are arbitrary. The two terms in
(\ref{eq83}) admit of a simple interpretation. They represent
damped-in-time waveforms traveling in the direction of
$x$-positive and $x$-negative.

In this connection it should be necessarily mentioned that for
$0<\alpha\leq 1$\,\, Eq. (\ref{eq82}) will describe an anomalous
diffusion with a potential. Then the function $\phi(x,t)$ becomes
a probability density. Thus the value has quite an other
interpretation \cite{19} in comparison with the above-mentioned
case.

\section{Interpretation and discussion}\label{sec:8}
Nature is permeated with oscillatory phenomena. Pulsating stars
and earthquakes, oscillating chemical reactions and long term
variations of the Earth's magnetic fields, circadian rhythms and
beats of the heart, electromagnetic waves and modes of oscillation
of the atom nucleus are examples of this kind of phenomena among
many others. Electrical and mechanical oscillators are everyday
constituents in the world of engineering. Usually an oscillator is
something that behaves cyclically, changing in some way, but
eventually getting back to where it started again. The fractional
calculus extends our representation about oscillatory phenomena.
The temporal evolution of fractional oscillator models occupies an
intermediate place between exponential relaxation and pure
harmonic oscillations. The fractional oscillator clearly
demonstrates a pure algebraic decay together with an exponentially
damped harmonic motion.

The modification of the conventional representation of the
Hamilton equations is conditioned on a random interaction of the
subsystems with environment. Each subsystem is governed by its own
internal clock. Although its dynamics is described by the ordinary
Hamiltonian equations,\,\,\, their coordinates and momenta depend
on the operational time. The passage from the operational time to
the physical time through the averaging procedure accounts for the
interaction of the subsystem with environment. Consequently, the
whole of subsystems behaves as a fractional system.\\

The author thanks D. Dreisigmeyer for fruitful discussions.

%

\end{document}